\documentclass[reprint, amsmath, amssymb, aps, showkeys, nolongbibliography, numerical]{revtex4-2}

\usepackage[usenames,dvipsnames]{xcolor}
\usepackage[mathlines]{lineno}

\usepackage{graphicx}
\usepackage[colorlinks=true, linkcolor=blue, allcolors=blue]{hyperref}
\usepackage[mathlines]{lineno}

\begin{document}

\title{Conformal invariance in out-of-equilibrium Bose-Einstein condensates governed by the Gross-Pitaevskii Equation}

\author{J.~Amette Estrada}
\affiliation{Universidad de Buenos Aires, Facultad de Ciencias Exactas y Naturales, Departamento de Física, Ciudad Universitaria, 1428 Buenos Aires, Argentina,}
\affiliation{CONICET - Universidad de Buenos Aires, Instituto de F\'{\i}sica Interdisciplinaria y Aplicada (INFINA), Ciudad Universitaria, 1428 Buenos Aires, Argentina.}
\author{M.~Noseda}
\affiliation{Universidad de Buenos Aires, Facultad de Ciencias Exactas y Naturales, Departamento de Física, Ciudad Universitaria, 1428 Buenos Aires, Argentina,}
\affiliation{CONICET - Universidad de Buenos Aires, Instituto de F\'{\i}sica Interdisciplinaria y Aplicada (INFINA), Ciudad Universitaria, 1428 Buenos Aires, Argentina.}
\author{P.J.~Cobelli}
\affiliation{Universidad de Buenos Aires, Facultad de Ciencias Exactas y Naturales, Departamento de Física, Ciudad Universitaria, 1428 Buenos Aires, Argentina,}
\affiliation{CONICET - Universidad de Buenos Aires, Instituto de F\'{\i}sica Interdisciplinaria y Aplicada (INFINA), Ciudad Universitaria, 1428 Buenos Aires, Argentina.}
\author{P.D.~Mininni}
\affiliation{Universidad de Buenos Aires, Facultad de Ciencias Exactas y Naturales, Departamento de Física, Ciudad Universitaria, 1428 Buenos Aires, Argentina,}
\affiliation{CONICET - Universidad de Buenos Aires, Instituto de F\'{\i}sica Interdisciplinaria y Aplicada (INFINA), Ciudad Universitaria, 1428 Buenos Aires, Argentina.}

\date{\today}

\begin{abstract}
We study density isolines in quantum turbulence under the Schramm-Loewner framework using direct numerical simulations of the truncated Gross-Pitaevskii equation, in both spherical and cylindrical traps with three-dimensional dynamics. Density isolines develop increasing complexity as turbulence matures. As the systems evolves towards a thermalized regime, it spontaneously develops conformal invariance. In contrast to other systems exhibiting conformal invariance, this system manifests it during the transition towards disorder rather than to self-organization. We discuss a link between this behavior in quantum turbulence and other 4-wave interacting systems.   
\end{abstract}


\maketitle

We lack a general theory for out-of-equilibrium, strongly interacting systems. Fluids, quantum chromodynamics, and condensed matter provide prime examples that showcase classical and quantum instances of strongly coupled many-body systems. Recent advances in the study of their out-of-equilibrium dynamics and scaling laws often involve scrutinizing their symmetries and possible universality classes \cite{Iyer_21}. An example is given by classical turbulence, where a symmetry broken by the presence of viscosity results in the  gradual restoration of symmetries in a statistical sense as nonlinear coupling, controlled by the Reynolds number, increases \cite{frisch_turbulence_1995}. 

Many-body quantum systems are another example of interest, driven by precisely controlled experiments, as well as by recent progresses in theoretical physics \cite{Padayasi_23}. In this context, atomic Bose-Einstein condensates (BECs) have provided experimental, theoretical, and numerical ways to explore out-of-equilibrium dynamics in an ample variety of configurations \cite{Henn2009, ClarkdiLeoni2015, ClarkdiLeoni2017, Shukla2019, makinen2023, AmetteEstrada2022, Barenghi2023, Tsubota2017}. The roads towards thermalization in BECs, passing through transient non-thermal fixed points \cite{Madeira_2022}, have been successfully explored using several finite temperature models. Of these, simulations using different formulations of the truncated or stochastic Gross-Pitaevskii equation (GPE) have shown good agreement with experimental results up to the condensate critical temperature \cite{Davis2006, Proukakis2008, Krstulovic2011b, AmetteEstrada2022_AVS}.

Recent advances linking out-of-equilibrium systems with quantum field theories have been made possible by the identification of conformal invariance. 
This is a stronger symmetry than scale invariance, as under this symmetry systems are 
invariant under transformations that preserve angles with rescaling that depends on position. In many cases these  advances were allowed by a precise connection between one-dimensional Brownian motion and two-dimensional conformal curves provided by Schramm-Loewner evolution (SLE) \cite{Schramm_2000}, which enabled direct examination of conformal invariance in numerical and experimental data. Noteworthy examples are given by two-dimensional classical \cite{bernard_conformal_2006} and quantum turbulence \cite{Panico_23}, surface quasigeostrophic turbulence \cite{bernard_inverse_2007}, and rotating turbulence \cite{thalabard_conformal_2011}. In these systems the out-of-equilibrium dynamics manifest as a self-similar inverse cascade, wherein energy injected at small scales spontaneously organizes into large scale patterns. Other applications of conformal invariance include percolation \cite{pose_shortest_2014} and condensed matter \cite{saberi_conformal_2008, Padayasi_23}.

In this letter we study the evolution of density isolines and their conformal invariance in quantum turbulence using direct numerical simulations of the truncated Gross-Pitaevskii equation (GPE). At zero temperature, dilute atomic BECs can be described by the GPE,
\begin{equation}
     i \hbar \frac{\partial \psi ({\bf x},t)}{\partial t} = \left[  -\frac{ \hbar ^2 }{2 m}\nabla^2 + g |\psi ({\bf x},t)|^2 + V({\bf x}) \right]\psi ({\bf x},t) ,
     \label{eq: GPE}
\end{equation}
where $m$ is the mass of the bosons, $g = 4 \pi a \hbar^2/m$, $a$ is the s-wave scattering length, $V({\bf x})$ is an external potential, and $\psi$ is the order parameter. When truncated to a finite number of excited modes using a projector operator $\mathbb{P}_K [\psi] = \sum_{|{\bf k}| \le K} \hat{\psi}_{\bf k}(t) e^{i {\bf k} \cdot {\bf x}}$, where $\hat{\psi}_{\bf k}$ are Fourier coefficients and ${\bf k}$ the wave vectors, Eq.~(\ref{eq: GPE}) results in the truncated GPE. When the truncated GPE is integrated for long times, it reaches a finite-temperature thermodynamic equilibrium consistent with the microcanonical ensemble given an initial energy, momentum, and number of particles \cite{Shukla2019}. Fluctuations in these states provide a classical description of thermal fluctuations by approximating the quantum field of highly occupied levels by a classical field, and were seen to agree with experimental results in many situations \cite{Proukakis2008}. 

\begin{figure*}
\centering
    \includegraphics[width=\textwidth]{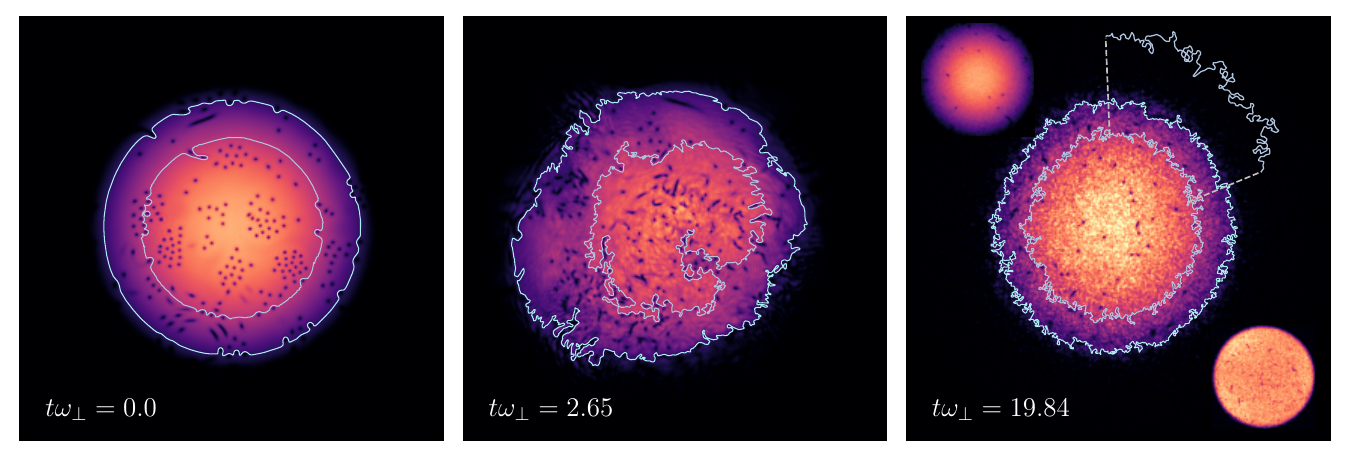}
    \caption{Mass density $\rho (x,y,z=0)$ at different times in the cigar harmonic potential; lighter colors correspond to larger densities. Inner and outer solid lines correspond respectively to density isolines with $\rho = 0.6 M/L^3$ and $\rho = 0.2 M/L^3 $. The complexity of the isolines grows with time. The right panel shows a zoom of a section of one isoline to illustrate complexity. The top left and bottom right insets in the right planel show respectively a thermal and a box trap state at the same time.}
    \label{Figure:densities}
\end{figure*}

We integrate this equation starting from out-of-equilibrium initial conditions, letting them freely evolve towards thermalization. The truncated GPE is solved with an axisymmetric cigar potential $V({\bf x}) = m \omega_\perp^2 (x^2+y^2)/2 $ (similar results were obtained with spherical traps, and with box traps that result in a more homogeneous mass distribution inside the trap; both cases are briefly discussed below). The system is integrated in a cubic domain of dimensions $[-\pi,\pi]L \times [-\pi,\pi]L \times [-\pi,\pi]L$, using a Fourier-based pseudo-spectral method with a spatial grid of $N^3 = 512^3$ points \cite{AmetteEstrada2022}. The $2/3$ rule is used to control aliasing instabilities, and a fourth-order Runge-Kutta method is used for time integration with the GHOST parallel code, which is publicly available \cite{Mininni2011}. To get initial conditions with minimum possible thermal excitations, we prepare the gas in a state with randomly distributed quantized vortices using the procedure described in \cite{Mller2020}, and then we integrate those conditions to a steady state using the Advective Real Ginzburg-Landau equation (ARGLE), which converges asymptotically to fixed (albeit not necessarily stable) points of GPE. ARGLE is dissipative and is obtained from GPE by means of a Wick rotation by which time becomes imaginary, plus a local Galilean transformation to impose an initial velocity field consistent with the presence of the quantized vortices \cite{Nore1997}. Evolving ARGLE results in an initial state with minimal phonon excitations, which is then integrated with the truncated GPE to obtain the results that follow. To perform comparisons we also generated thermal states (i.e., states dominated by phonons) using the procedure described in \cite{Shukla2019, Amette_2024}. These states are in thermal equilibrium, and have no vortices except for those randomly excited by the thermal fluctuations.

All results are shown in units of a characteristic speed $U$, the unit length $L$ (proportional to the condensate mean radius), and a unit total mass $M$. All parameters in Eq.~(\ref{eq: GPE}) can be fixed by setting the speed of sound as $c = (g \rho_0/m)^{1/2} = 2U $, the condensate healing length as $\xi = \hbar/(2m \rho_0 g)^{1/2} = 0.0088 L$, the trapping frequency to $\omega_\perp = 35 \, U/L$, and the central density as $\rho_0 = 1 M/L^3$. Quantities can then be scaled by setting dimensional values for $U$, $L$, and $M$. In experiments typical dimensional values are $L\approx 10^{-4}$ m and $c \approx 2 \times 10^{-3}$ m/s \cite{White2014}. This results in $\xi \approx 1.4 \times 10^{-7}$ m and a trap frequency $\omega_\perp \approx 26$ Hz. For $10^7$ $^{87}$Rb atoms, particle densities of $\approx 10^{13}$ cm$^{-3}$ atoms are compatible with experiments \cite{Henn2008}. 

Figure \ref{Figure:densities} shows slices of the density $\rho(x,y,0)$ at different times, as well as two density isolines with $\rho = 0.2 M/L^3$ and $\rho = 0.6 M/L^3$. The mass density is obtained using Madelung's transformation, $\psi (\mathbf{x},t) = [\rho(\mathbf{x},t)/m]^{1/2} e^{i S(\mathbf{x},t)}$, where $\rho(\mathbf{x},t)$ is the condensate mass density, and its pointwise velocity is $\mathbf{v} = (\hbar/m) \boldsymbol{\nabla} S (\mathbf{x},t)$ \cite{Nore1997}. The evolution of $\rho$ and of its isolines provides a first glimpse at how complexity develops. The initial condition is smooth, with randomly placed quantized vortices (seen as regions with low mass density). At intermediate times ($t \omega_{\perp} = 2.65$) the system becomes turbulent (i.e., in a transient non-thermal fixed point), and strong fluctuations can be seen in $\rho$ accompanied by large-scale deformations of the condensate. Finally, at late times ($t \omega_{\perp} = 19.84$) the condensate recovers some isotropy, but isolines exhibit their highest complexity. Isolines at different values of $\rho$ become similar, and as will be shown below, a non-negligible fraction of the fluctuations correspond to phonons resulting from the decay of turbulence. Indeed, from $t \omega_{\perp} \ge 7$ isolines remain qualitatively similar to those shown at $t \omega_{\perp} = 19.84$ in Fig.~\ref{Figure:densities}.

To quantify fractality and conformal invariance of density isolines in this system, we build an ensemble of isolines with different densities ranging from $\rho = 0.2 M/L^3$ to $0.6 M/L^3$. In the cylindrical traps, using the translational symmetry and to increase statistics, we extract curves every $0.12 L$ in the $z$ direction at each time (even though both cylindrical and spherical traps are observed to be consistent with conformal invariance, in the following we show results from the former except when explicitly noted, as a result of its convenience to extract more curves to improve the statistics). Note that all isolines are closed. To work with open curves we set the origin at some arbitrary point, and we extract curves with at least 200 points in a direction given by the rule that larger mass densities must be to the left. In the cylindrical traps, this procedure resulted in a total of 976 curves with an average of 1519 points per curve.

\begin{figure}
\centering
    \includegraphics[width=1.02\columnwidth, trim=0.55cm 0.55cm 0.55cm 0.55cm,clip]{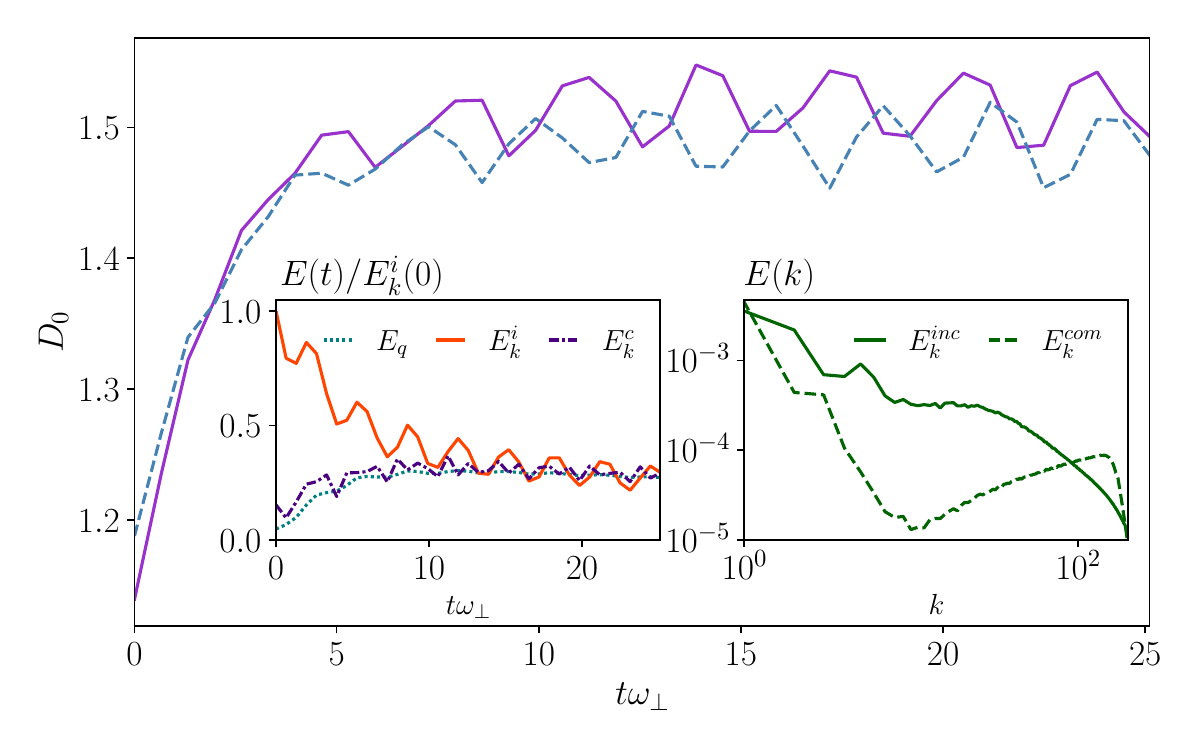}
    \caption{Time evolution of the fractal dimension $D_0$ of $\rho$-isolines in the cylindrical (solid purple line) and in the spherical harmonic traps (dashed blue line). The left inset shows the time evolution of different energy components in the cylindrical harmonic trap, normalized by the initial incompressible kinetic energy (see text for references).  Oscilations are in phase in all quantities and correspond to the condensate breathing mode. The right inset shows compressible and incompressible kinetic energy spectra after $D_0$ stabilizes.}
    \label{Figure:Fractal_dimension}
\end{figure}

The fractal dimension $D_0$ gives a first estimation of the self-similarity and complexity of the isolines. We estimated it using the yardstick method as described in \cite{Mandelbrot1983, giordanelli_conformal_2016}. Similar results were obtained using the box counting method. Figure \ref{Figure:Fractal_dimension} shows $D_0$ for the ensemble of curves as a function of time, for the cylindrical and spherical harmonic traps. For the cylindrical trap, the left inset in Fig.~\ref{Figure:Fractal_dimension} shows the time evolution of the condensate's quantum energy $E_q= \hbar^2/(2m) \langle (\nabla \sqrt{\rho})^2 \rangle$, and kinetic energy $E_k = \langle \rho v^2 \rangle/2$; the latter being further decomposed into incompressible kinetic energy $E_k^i$ and compressible kinetic energy $E_k^c$ using a Helmholtz decomposition \cite{Nore1997}. This allows us to quantify the kinetic energy in turbulent motions ($E_k^i$), and in sound and thermal excitations ($E_k^c$). The right inset in Fig.~\ref{Figure:Fractal_dimension} shows the spatial spectra associated to these two energies.

In Fig.~\ref{Figure:Fractal_dimension} we can identify evolution stages similar to those observed in Fig.~\ref{Figure:densities}. At $t=0$ the fractal dimension of the curves is close to $1$, and the compressible and quantum energies are significantly smaller than the incompressible kinetic energy. As this energy component decays, turbulence develops and strong fluctuations grow (as evidenced by the growth of $E_k^c$ and $E_q$). The compressible kinetic energy grows as a result of the excitation of sound waves in the system. During this transient $D_0$ also grows in time, until for $t \omega_\perp \gtrsim 3.5$, $D_0$ stops growing and the energy components approach equipartition, i.e., $E_k^i \approx E_k^c \approx E_q$. Beyond this stage, the oscillations seen in $D_0$ and in the energies are associated to the breathing mode of the condensate in the trap. In the cylindrical trap the fractal dimension stabilizes at a mean value of $D_0 \approx 1.51$, and it increases or decreases around this value as the cloud compresses of expands with the breathing mode oscillations. The turbulent direct cascade of energy and the resulting excitation of small-scale fluctuations (i.e., density fluctuations and sound waves) are required for the isolines to become fractal. Note in the inset of Fig.~\ref{Figure:Fractal_dimension} that broad incompressible and compressible kinetic energy spectra develop once $D_0$ stabilizes.

\begin{figure}
\centering
    \includegraphics[width=\columnwidth]{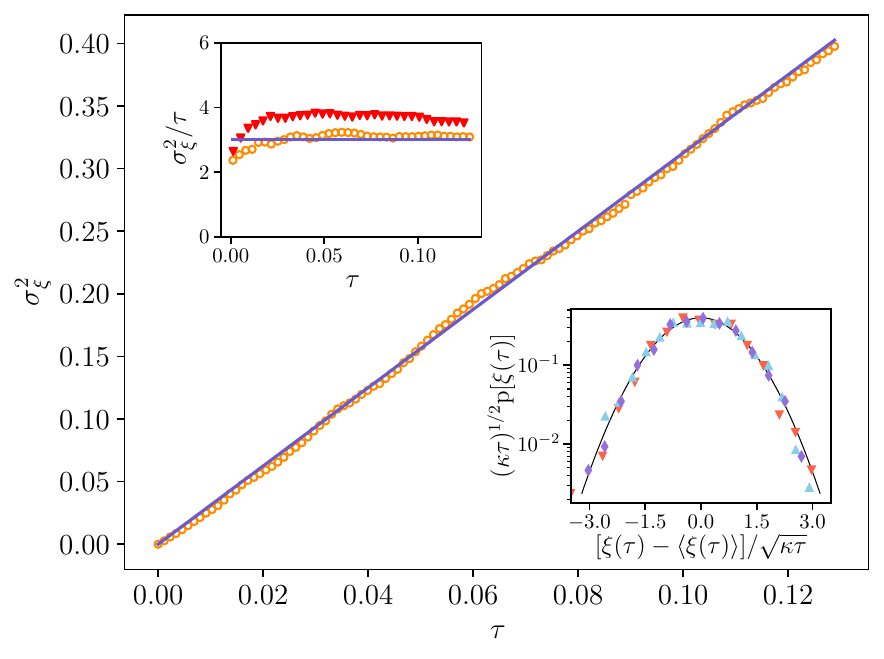}
    \caption{Scaling of the variance of the ensemble of driving functions with the Loewner time $\tau$, with a fit with $\kappa = 3.12 \pm 0.18$. The bottom inset shows the histogram of the renormalised drivings at three different times $\tau = 0.015$, $0.065$, and $0.13$, with a Gaussian in the solid black line as a reference. The top inset shows the variance compensated by $\tau$, and the horizontal line indicates $\kappa = 3$. Results for thermal states are shown in red triangles and have $\kappa = 3.76 \pm 0.26$.}
    \label{Fig:Direct_test}
\end{figure}

Is the fractal dimension a manifestation of a more general symmetry in the system? To answer this question we now focus on the cylindrical trap, as we can extract more curves and obtain better statistics from that geometry. To test the system for conformal invariance we want to statistically associate the ensemble of isolines of $\rho$ to a family of driving functions in SLE. Such a family must satisfy the Loewner equation, which was discovered by Loewner as a way to describe the growth of a trace $\gamma$ in the complex domain. To work with chordal traces, which are curves that start at the origin and grow towards infinity and are limited to half of the complex plane, we use a holomorphic transformation of the isolines. In other words, each isoline is described as a sequence of points in the complex plane $\{z_0, z_1, \dots, z_N \}$, where $z_0$ is set to the origin. To convert them into chordal traces we apply the M\"obius transformation, $\zeta_i = z_N z_i (z_i - z_N)^{-1}$, as done in \cite{saberi_conformal_2008, thalabard_conformal_2011}. For the resulting traces the Loewner equation is $\partial_{\tau} g_{\tau} (\zeta) = 2  [g_\tau(\zeta) - \xi(\tau)]^{-1}$, where $\tau$ is the Loewner time (not to be confused with the physical time $t$) that parameterizes the evolution of the trace, $g_\tau(\zeta)$ is a conformal transformation that maps the trace in the half-plane into the real axis, and $\xi(\tau)$ is the so-called driving function, which is an unknown one-dimensional real continuous and stochastic function that encodes the trace $\gamma$. If $\xi(\tau)$ is Gaussian and corresponds to Brownian motion, then the isolines are conformal invariant. Moreover, under this conditions the variance of $\xi$ over all traces has a diffusivity $\kappa$ such that $\sigma_\xi^2 = \left< [\xi (\tau) - \left< \xi (\tau)\right>]^2  \right> = \kappa \tau$. The value of $\kappa$ allows quantification of the complexity of  isolines, and classification of physical phenomena into universality classes. This enables powerful associations between different physical systems, as in the case of boundaries of clusters in Ising models near the critical point \cite{chelkak_convergence_2014} and zero-vorticity isolines in two-dimensional turbulence \cite{bernard_conformal_2006}.

\begin{figure}
\centering
    \includegraphics[width=\columnwidth]{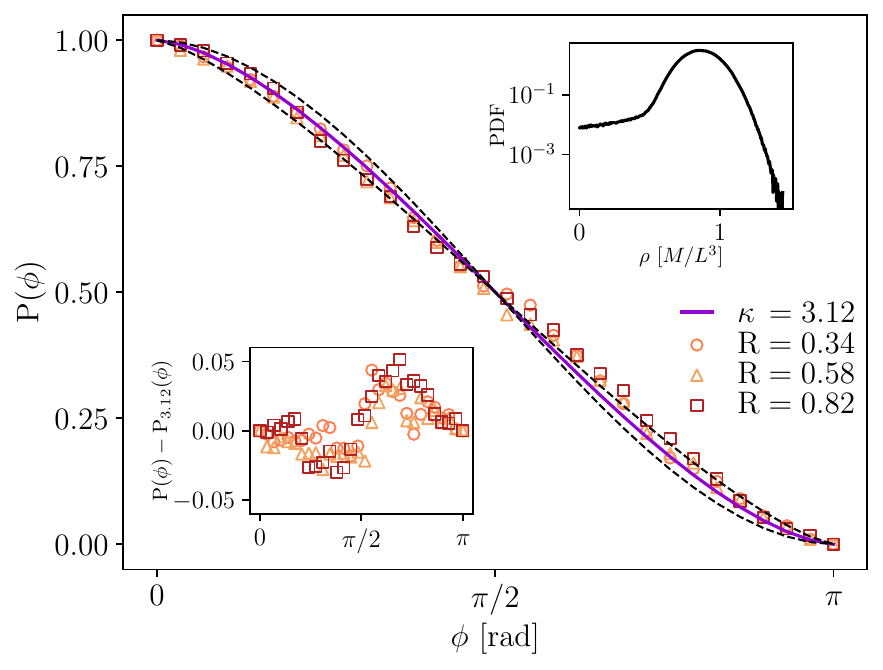}
    \caption{Left passage probability $P_\kappa(\phi)$ of traces passing to the left of points with different $R$, in the cylindrical harmonic trap. The violet curve shows the theoretical prediction with $\kappa = 3.12$. Black dashed lines indicate $10 \%$ envelopes in the value of $\kappa$. The bottom inset shows residues. The top inset shows the probability density function of mass density in the same trap, note the deviations from Gaussianity.}
    \label{Fig:left-passage-probability}
\end{figure}

To obtain the driving functions we use the zipper algorithm with vertical slits \cite{marshall_convergence_2007, Kennedy2007}. This algorithm gradually wraps the half-plane traces into the real axis by using a composition of transformations. The composition of these transformations gives us $g_\tau(\zeta)$, allowing computation of $\xi(\tau)$. For convenience the final Loewner times are remapped to one, using the scaling property of SLE \cite{lawler_introduction_nodate}. As already mentioned, if the traces effectively result from Brownian driving functions under SLE, the ensemble of $\xi(\tau)$ should converge statistically to a Gaussian process with variance $\kappa \tau$. To verify this we first use the Kolmogorov-Smirnov test. For dynamical times $t \omega_\perp \ge 3.5$ (i.e., once the fractal dimension of the isolines stabilizes) the test is passed, while for $t \omega_\perp < 3.5$ the hypothesis is rejected. This indicates that the system evolution towards a thermalized state is important to obtain conformal invariance. Figure \ref{Fig:Direct_test} shows a direct confirmation of the invariance in the cylindrical harmonic trap, displaying the variance $\sigma_\xi^2$ as a function of the Loewner time for the ensemble of driving functions with $t \omega_\perp \ge 3.5$. The variance is in good agreement with $\kappa \tau$ scaling, with $\kappa = 3.12 \pm 0.18$ obtained from a linear best fit. Insets show $\sigma_\xi^2$ compensated by $\tau$, and probability density functions of the driving functions at different Loewner times compared against a Gaussian distribution. It is worth noting that isolines in the spherical trap and in the box trap yield similar results. However, the purely thermal states result in a larger value of $\kappa$, with a linear best fit yielding $\kappa = 3.76 \pm 0.26$ (see the inset in Fig.~\ref{Fig:Direct_test}).

A direct relation exists between $\kappa$ and the fractal dimension of SLE curves, $D_0 = \textrm{min}\{ 1+\kappa/8, 2\}$ \cite{beffara_dimension_2008}. In the cylindrical harmonic trap $\kappa = 3.12$ results in $D_0 \approx 1.4$, which is compatible with the mean fractal dimension in Fig.~\ref{Figure:Fractal_dimension} for $t \omega_\perp \ge 7$, specially considering that the direct estimation of $D_0$ has larger uncertainties.

Finally, as a last test of conformal invariance, we compute the left passage probability for the ensamble of traces. This property quantifies how often each trace leaves an arbitrary point in the complex plane $\zeta_* = R e^{i \phi}$ to its left, where $R$ is the distance of the point to the origin and $\phi$ its angle. Schramm obtained a theoretical expression for this probability if the traces satisfy SLE, that depends solely on $\kappa$ and $\phi$ \cite{schramm_percolation_2001},%
\begin{equation}
    P_\kappa(\phi) = \frac{1}{2} + \frac{\Gamma\left(\frac{4}{\kappa}\right)}{\sqrt{\pi}\: \Gamma\left(\frac{8-\kappa}{2\kappa}\right)} \ _2F_1 \left(\frac{1}{2}, \frac{4}{\kappa}, \frac{3}{2}, -\cot^2 \phi \right) \cot(\phi),
    \label{eq:schramm}
\end{equation}
where $\Gamma$ is the Gamma function and $_2F_1$ is the Gauss hypergeometric function. Figure \ref{Fig:left-passage-probability} shows the comparison between Eq.~(\ref{eq:schramm}) for $\kappa = 3.12$ and $\phi \in [0,\pi]$, and the results for our traces at three values of $R$ in the range of Loewner times in which scaling of the variance of the driving functions is linear (from $\tau = 0.03$ to $0.13$).

We showed that out-of-equilibrum BECs described by the Gross-Pitaevskii equation evolve towards thermalization through a conformal invariant transient non-thermal fixed point. Unlike previous studies \cite{bernard_conformal_2006, bernard_inverse_2007, thalabard_conformal_2011, Panico_23}, here SLE behavior is obtained: (1) For all isolines of the density (instead of, e.g., one specific value of the vorticity \cite{bernard_conformal_2006}). (2) As a result of the time evolution, with the dynamics of the system modulating the fractal dimension of the curves. (3) In a system that evolves towards thermalization with a direct energy cascade (as opposed to self-organized systems with inverse cascades \cite{bernard_conformal_2006, bernard_inverse_2007, Panico_23}).
For very long times, in the final thermalized state, we can expect $\kappa$ to asymptotically approach 4, as indicated by our analysis of thermal states, and as suggested by exact results for discrete Gaussian free fields \cite{Schramm_2009}. The possibility of the system reaching this value is also of interest as $\kappa \approx 4$ was also observed in surface quasigeostrophic turbulence \cite{bernard_inverse_2007, falkovich_conformal_2007}, albeit the statistics in that system is not Gaussian, and the two-point correlation function deviates from that expected for Gaussian free fields (as is also the case in mass fluctuations in GPE \cite{Clark_di_Leoni_2018}). Moreover, our results indicate that the transient turbulent regime has a smaller $\kappa$ closer to 3 (a value of $\kappa=6$, for random wavefunctions in the semi-classical limit \cite{Bogomolny_2007}, is also discarded by our results). The differences are consistent with the fact that the transient out-of-equilibrium state is interacting, and displays deviations from Gaussianity caused by the presence of quantized vortices (see Fig.~\ref{Fig:left-passage-probability} and Ref.~\cite{Paoletti_2008}). Interestingly, a value of $\kappa = 3$ is obtained for domain walls in the critical Ising model, and $\kappa = 2.88 \pm 0.08$ was found in surface wave experiments \cite{Noseda24}. The latter system has a direct connection with the dynamics described by GPE.

GPE describes a larger set of systems with 4-mode interactions. On the one hand, GPE can be obtained from the Hamiltonian $H=\int [\hbar^2 |\boldsymbol{\nabla} \psi|^2/(2m) + g |\psi|^4/2] d^3 x$, thus corresponding to a general equation for non-relativistic scalar fields with $|\psi|^4$ interaction. On the other hand, for a nearly monochromatic wave package centered around wave vector ${\bf k}_0$ and frequency $\omega_0$, the 4-wave interaction Hamiltonian $H=\int \omega_{\bf k} |a_{\bf k}|^2 d^3k + \int T_{\bf 123k} a_{\bf k}^* a_{\bf 1}^* a_{\bf 2} a_{\bf 3} \delta({\bf k} + {\bf k}_1 - {\bf k_2} - {\bf k}_3) d^3k_1 d^3k_2 d^3k_3 d^3k$ (where $a_{\bf k}$ is the amplitude of the wave with wave vector ${\bf k}$, and $T_{\bf{123k}}$ is the scattering amplitude of four waves), results in the general equation of motion $i \dot{a}_{\bf k} = \omega_{\bf k} a_{\bf k} + \int T_{\bf 123k} a_{\bf 1}^* a_{\bf 2} a_{\bf 3} \delta({\bf k} + {\bf k}_1 - {\bf k_2} - {\bf k}_3) d^3k_1 d^3k_2 d^3k_3$. This equation in a homogeneous medium and for $a_{{\bf k}_0 + {\bf q}} = \psi_{\bf q} e^{i \omega_0 t}$ is equivalent to GPE \cite{falkovich_2011}. As a result, we can expect other systems with interaction of four waves or normal modes to display conformal invariance (experimental evidence of this symmetry has been recently reported for gravity wave turbulence \cite{Noseda24}, with a value of $\kappa$ compatible with the value reported here within error bars). Finally, GPE can be rewritten using Madelung's transformation as the Euler equation of a classical and compressible barotropic fluid (albeit with a specific equation of state) \cite{Nore1997},  suggesting that conformal invariance could be obtained in some regimes of compressible turbulent flows, in agreement with was recently found for weakly compressible two-dimensional turbulence \cite{Puggioni_2020}.

In recent years, connections between field theory and turbulence have provided new insights into out-of-equilibrium dynamics \cite{Migdal_94, bernard_conformal_2006, Iyer_21, Muller_21}. For systems having an underlying conformal structure, many tools from field theory can be used to study their scale invariance (as well as their deviations). The connections discussed here between GPE and other 4-wave interacting systems can have applications in many other physical systems, providing a possible general link between conformal invariance and multifractal scaling in systems in which the long wavelength dynamics is governed by nonlinear interactions between four modes.

\begin{acknowledgments}
The authors acknowledge financial support from UBACyT Grant No.~20020170100508BA and PICT Grant No.~2018-4298. J.A.E.~and M.N.~contributed equally to this work.
\end{acknowledgments}

\bibliography{ms}

\end{document}